\documentclass[epj]{webofc}
\usepackage[varg]{txfonts}   
\usepackage{graphicx,color} 
\usepackage{bm}       
\usepackage{amsmath}  
\usepackage{amssymb}
\usepackage{dsfont}
\usepackage{multirow}
\newcommand{\ntwolo}{N$^2$LO}
\newcommand{\nnn}{3N}
\newcommand{\nalpha}{$n$-$\alpha$}
\woctitle{21st International Conference on Few-Body Problems in Physics}
\begin{document}
\title{Green's Function Monte Carlo Calculations with Two- and
Three-Nucleon Interactions from Chiral Effective Field Theory}
\author{J.~E.~Lynn\inst{1}\fnsep\thanks{\email{joel.lynn@gmail.com}}
}

\institute{Theoretical Division, Los Alamos National Laboratory, Los
Alamos, New Mexico 87545, USA
          }

\abstract{
I discuss our recent work on Green's function Monte Carlo (GFMC)
calculations of light nuclei using local nucleon-nucleon interactions
derived from chiral effective field theory (EFT) up to
next-to-next-to-leading order (\ntwolo{}).
I present the natural extension of this work to include the consistent
three-nucleon (3N) forces at the same order in the chiral expansion.
I discuss our choice of observables to fit the two low-energy constants
which enter in the 3N sector at \ntwolo{} and present some results for
light nuclei.
}
\maketitle
\section{Introduction}
\label{intro}
Accurate predictions of properties of light nuclei require at least two
ingredients: (1) validated nuclear interactions and (2) accurate
many-body computational methods.
While quantum Chromodynamics is known to be the ultimate source of the
first, arguably the best current path to validated nuclear interactions
is through chiral effective field theory
(EFT)~\cite{ref:epelbaum2009,ref:machleidt2011}.
The second may come in many forms -- hyperspherical harmonics, the no-core
shell model, and Green's function Monte Carlo (GFMC) to name a few --
but among the most accurate of these is the GFMC method.
Recently we have demonstrated the powerful effect of combining these two
ingredients (which was not possible before)~\cite{ref:lynn2014} by
studying properties of the lightest nuclei $A=3,4$ with chiral
interactions up to \ntwolo{}, neglecting however, the corresponding
three-nucleon (3N) interaction which enters at \ntwolo{}.
Now we include the consistent 3N interaction arising at \ntwolo{} in the
chiral expansion, fitting the two low-energy constants which enter at
this order to the $^4$He binding energy and to \nalpha{} $P$-wave
elastic scattering phase shifts~\cite{ref:lynn2015}.

\section{The Fits}
\label{sec:fits}



\begin{figure}[h]
\centering
\includegraphics[width=7cm,clip]{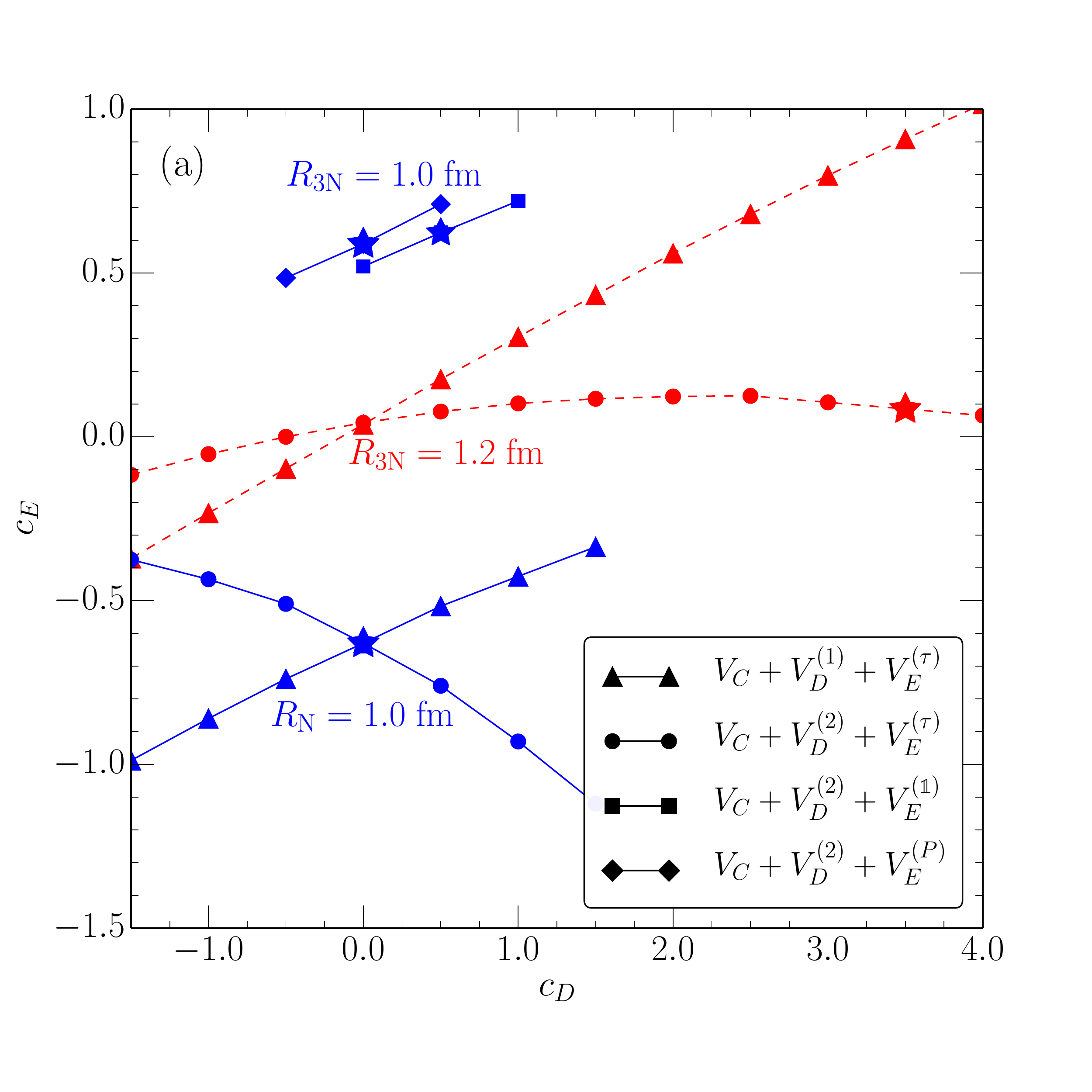}
\includegraphics[width=7cm,clip]{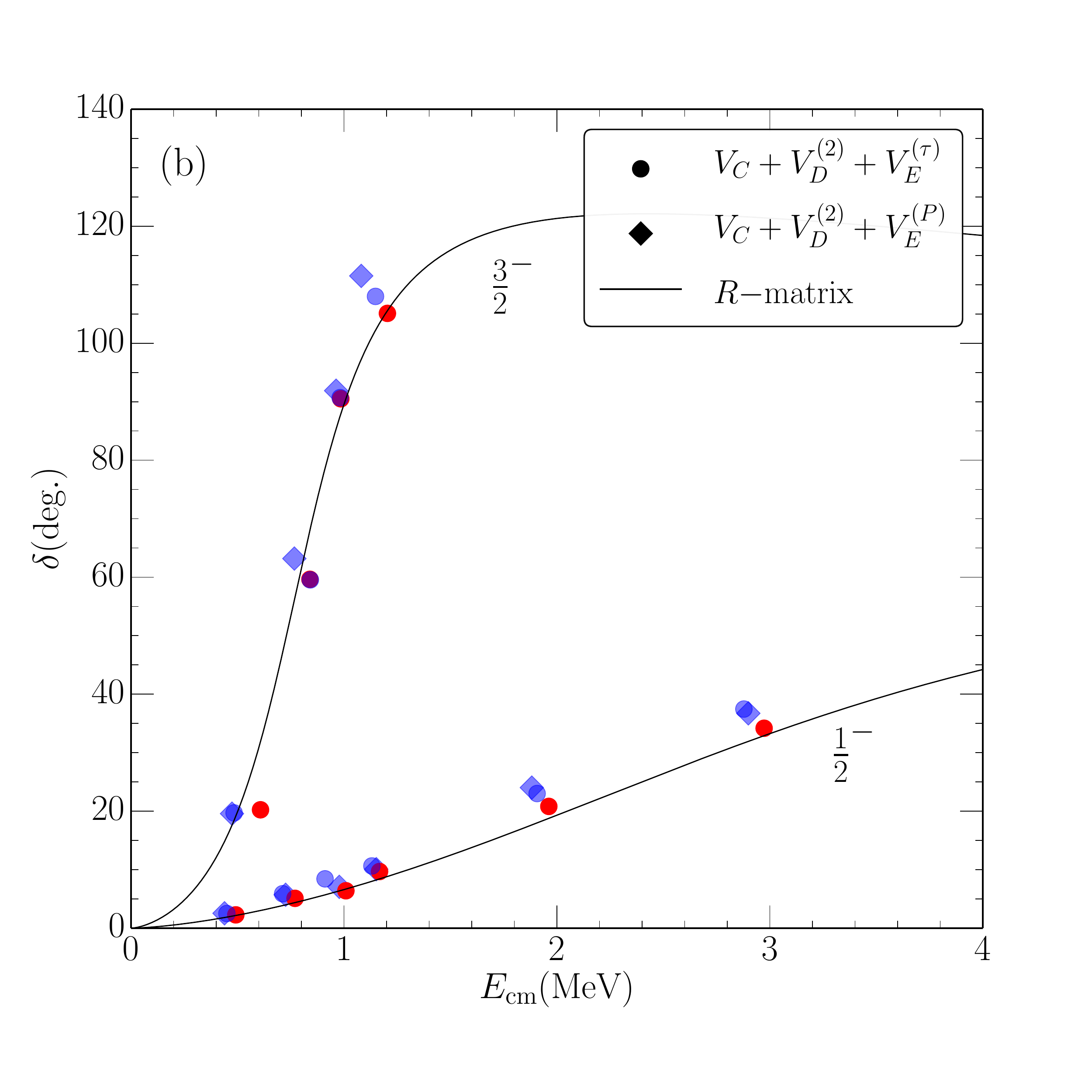}
\caption{
   Fitting $c_D$ and $c_E$.
   The GFMC statistical uncertainties are the size of the symbols or
   smaller.
   Blue (Red) symbols correspond to $R_\text{\nnn{}}=1.0$~fm
   ($R_\text{\nnn{}}=1.2$~fm), with $R_\text{\nnn{}}$ a 3N cutoff
   parameter.
   (a) Curves of $c_E$~vs.\!~$c_D$ obtained by fitting the $^4$He
   binding energy.
   The fits were obtained at the points:
   The lines are a guide to the eye.
   The stars correspond to the values of $c_D$ and $c_E$ which fit
   both the $^4$He binding energy and the \nalpha{} $P$-wave
   phase shifts.
   No fit to both observables can be obtained for the case with
   $R_\text{\nnn{}}=1.2$~fm and $V_D^{(1)}$.
   See Ref.~\cite{ref:lynn2015} for details.
   (b) $P$-wave \nalpha{} elastic scattering phase shifts compared with
   an $R$-matrix analysis of the data.}
\label{fig:fits}
\end{figure}

We fit the two low-energy constants $c_D$ and $c_E$ to $^4$He binding
energy Fig.~\ref{fig:fits}a and to $P$-wave \nalpha{} scattering phase
shifts~Fig.~\ref{fig:fits}b.
Details about the various 3N interactions $V_C+V_D^{(i)}+V_E^{(j)}$,
with $i=1,2,3$ and $j=\tau,\mathds{1},P$, which amount to different
operator choices for the terms proportional to $c_D$ and $c_E$ are
available in~\cite{ref:lynn2015}.
It is significant that chiral interactions at \ntwolo{} have sufficient
freedom to fit both properties of light nuclei~(Fig.~\ref{fig:fits}a and
Table~1) and the splitting of the $P$-wave \nalpha{} scattering phase
shifts (Fig.~\ref{fig:fits}b).
See~\cite{ref:lynn2015} for more details.

\begin{table}
\label{tab:ais3}
\centering
\caption{
Comparison of results for $A=3$ in this work with
experimental values.
The values in parenthesis are the GFMC statistical errors.}
{\renewcommand{\arraystretch}{1.00}
\vspace{2ex}
\begin{tabular}{cccccc}
&\multirow{2}{*}{$R_0$ (fm)}&\multicolumn{2}{c}{$E_B$ (MeV)}&
\multicolumn{2}{c}{$\sqrt{\langle r_\text{pt}^2\rangle}$ (fm)}\\
&&GFMC&Exp.&GFMC&Exp.\\
\hline
\multirow{2}{*}{$^3$H}&1.0&-8.33(1)&\multirow{2}{*}{-8.48}&1.55(3)&
\multirow{2}{*}{1.59}\\
&1.2&-8.35(4)&&1.55(4)&\\
\multirow{2}{*}{$^3$He}&1.0&-7.66(1)&\multirow{2}{*}{-7.72}&1.77(2)&
\multirow{2}{*}{1.76}\\
&1.2&-7.63(4)&&1.77(1)&\\
\hline
\end{tabular}}
\end{table}

\begin{acknowledgement}
This work was supported the U.S. DOE, Office of Nuclear Physics and by
the NUCLEI SciDAC program and used resources of NERSC, which is
supported by the U.S. DOE under Contract No.~DE-AC02-05CH11231.
\end{acknowledgement}

\end{document}